# Inertia-Aware Optimal Power Flow Using PINN in IBR-Dominated Power Systems

Mahyar Tofighi-Milani
*School of Electrical Engineering and Automation*
*Aalto University*
Espoo, Finland
Seyyed.tofighimilani@aalto.fi

Sajjad Fattaheian-Dehkordi
*School of Engineering*
*EPFL University*
Lausanne, Switzerland
sajjad.fattaheiandehkordi@epfl.ch

Franz Martin Rohrhofer
*Know-Center Research GmbH*
Graz, Austria
frohrhofer@know-center.at

Matti Lehtonen
*School of Electrical Engineering and Automation*
*Aalto University*
Espoo, Finland
matti.lehtonen@aalto.fi

***Abstract*—The problem of Optimal Power Flow (OPF) is central to the secure and economic operation of modern power systems. However, increasing renewable energy penetration, and decreasing system inertia pose significant challenges to conventional optimization-based OPF solvers. While machine learning approaches have demonstrated substantial computational speed-ups, purely data-driven methods often suffer from data dependency, limited generalization, and lack of guaranteed physical feasibility. This paper suggests a physics-informed neural network (PINN) framework for solving the OPF problem in renewable energy–dominated, low-inertia power systems. In contrast to conventional OPF formulations, the model explicitly incorporates a location-aware inertia constraint based on the concept of system inertia strength, which accounts for the electrical distance between generation units and disturbance locations. Simulation results on a 6 GW test system demonstrate high accuracy. The mean absolute error (MAE) for both the training and testing datasets is approximately 0.045% of the total system capacity. The findings demonstrate that the proposed PINN framework is capable of producing highly accurate OPF solutions while ensuring compliance with both physical laws and inertia-related constraints. Overall, the findings highlight the potential of physics-informed learning to enable secure, efficient, and computationally scalable OPF for future low-inertia power systems.**

## NOMENCLATURE

|  |  |
|---|---|
| $C_{con,i}$ / $C_{res,i}$ / $C_{res,i}^{HR}$ | The cost of $i^{th}$ conventional unit / RES unit / RES headroom reserve |
| $\Im$ / $\Re$ / $\Gamma$ | The set of conventional units / RES units / lines |
| $P_{con,i}$ / $P_{res,i}$ | Power generation of $i^{th}$ conventional / RES unit |
| $\alpha_i$ , $\beta_i$ , $\gamma_i$ | The cost coefficients of $i^{th}$ conventional unit |
| $\kappa_i$ , $\kappa_i^{HR}$ | Cost of $i^{th}$ RES and its headroom reserve |
| $P_{con,i}^{\min}$ / $P_{con,i}^{\max}$ | Minimum / maximum limits of $i^{th}$ conventional unit |
| $P_{res,i}^{\min}$ / $P_{res,i}^{\max}$ | Minimum / maximum limits of $i^{th}$ RES unit |
| $\overline{B}, B_{ij}$ | Susceptance matrix of the system and its (*i*,*j*) element |
| $\overline{\theta}, \theta_i$ | Angle vector of the system and its $i^{th}$ element |
| $\overline{P_G}$ / $\overline{P_D}$ | Power generation / demand vector |
| $P_{G,i}$ | $i^{th}$ element of $\overline{P_G}$ |
| $F_l$ , $F_l^{\max}$ | Power flow of line *l* and its maximum capacity |
| $l_1$ , $l_2$ | Beginning and end bus of line *l* |
| $H_{str}$, $H_{str}^{\min}$ | Strength of system inertia and its minimum limit |
| $\overline{H_G}$ / $\overline{H_B}$ | Inertia vector of generation / non-generation buses |
| $\overline{I_{H_B}}$ / $\overline{I_{P_D}}$ | Ones vector in the same dimension of $H_B$ / $P_D$ |
| $\overline{W}$ | Bus weighting matrix |
| $H_i^{HR}$ | Virtual inertia from $i^{th}$ RES's headroom reserve |
| $E_i^{HR}$ | Headroom reserved energy of $i^{th}$ RES |
| $\tau_i^{HR}$ | The supporting time of $i^{th}$ RES in the case of imbalance occurrence |
| $\overline{\hat{P}_G}^{(j)}$ / $\overline{P_G^*}^{(j)}$ | The vector of predicted generations / real generations |
| $N$ | Number of training data |

## I. INTRODUCTION

The Optimal Power Flow (OPF) problem plays a fundamental role in the secure and economic operation of modern power systems. It determines the optimal operating point of generation and network variables while satisfying physical laws and operational limits, enabling the objective of cost minimization [1], [2]. However, OPF is inherently nonlinear, and its complexity continues to grow with the expansion of network size, increasing penetration of renewable energy sources (RESs), and the introduction of new operational and security constraints in contemporary grids [1]. As a result, conventional solution approaches based on mathematical programming and iterative numerical solvers often face significant computational challenges, particularly for large-scale systems and time-critical applications.

Recently, machine learning (ML) and deep learning (DL) techniques have emerged as promising alternatives to accelerate OPF computations by shifting the computational burden from online optimization to offline training [3], [4], [5]. Approaches such as fully connected neural networks (FCNN), convolutional neural networks (CNN), and graph neural networks (GNN) have been investigated to predict optimal generator set-points or active constraint sets [3]. Methods such as Deep OPF have demonstrated considerable speed improvements compared to conventional solvers while incorporating reconstruction procedures to enforce equality constraints [4]. Data-driven ANN approaches have also been applied to emulate power flow and OPF solutions with substantial computational speed gains over traditional solvers [2].

Despite these promising developments, purely data-driven approaches face important limitations. They generally depend on substantial amounts of high-quality labeled data, they may produce physically infeasible solutions, and they often lack strong guarantees on constraint satisfaction and generalization performance [6]. In order for addressing these challenges, PINNs have recently attracted significant attention in power system applications. By incorporating physical laws directly into the learning procedure, PINNs reduce data dependency, improve accuracy, enhance computational efficiency while improving physical consistency and interpretability [6], [7]. Physics-informed reinforcement learning has also been implemented to enhance stability and security in real-time OPF [8].

Several pioneering studies have demonstrated the effectiveness of PINNs in power system problems. For example, a physics-informed framework for OPF has been proposed by incorporating the power flow equations into the learning process and providing guarantees on constraint violations in [9]. For power flow analysis, unsupervised PINNs (UPINNs) have been developed to solve power flow equations without relying on labeled data, achieving strong generalization capabilities [10].

Physics-informed typed GNNs have been proposed for OPF, enabling component-aware modeling without labeled data and achieving fast, accurate solutions [11]. Physics-informed frameworks have also been introduced for state estimation and inertia identification, with substantial computational speed-ups [12], [13]. Hierarchical PINN-based OPF frameworks have also been developed for hybrid AC/DC systems to improve coordination and computational efficiency [14]. Physics-informed learning with feasibility correction has further improved OPF optimality and convergence performance [15] . Recent investigations further highlight the scalability and efficiency of PINNs in transient stability analysis compared to conventional numerical solvers [16]. Decentralized multi-branch PINNs have demonstrated scalable and feasible OPF solutions [17].

While most ML-assisted OPF and PINN-based OPF formulations concentrate on steady-state operational constraints, modern power systems are increasingly challenged by dynamic stability issues. The growing penetration of RESs and converter-based resources has significantly reduced system inertia, leading to faster frequency dynamics and increased vulnerability to disturbances [18]. Inverter-based resource (IBR)-dominated power systems are characterized by higher rates of change of frequency (RoCoF) and deeper frequency nadirs, which may trigger load shedding or even cascading failures [19]. Furthermore, the increasing integration of converter-based technologies such as clean-hydrogen electrolysers and the deployment of grid-forming control strategies highlight the critical role of virtual inertia in maintaining grid stability [20], [21].

To mitigate declining inertia, various technical and market-based solutions have been proposed. These include virtual inertia emulation through power electronic converters [18], optimal procurement of virtual inertia and frequency reserve services [22], [23], and dynamic support from advanced electrolyzer technologies capable of providing fast frequency response [24]. Consequently, virtual inertia is anticipated to become a key component in ensuring frequency stability in future IBR-dominated power systems.

Nevertheless, inertia is rarely incorporated explicitly into OPF formulations beyond simplified security margins. In most existing OPF frameworks—whether conventional or ML-based—the problem is treated predominantly as a static optimization task, neglecting the minimum inertia required to ensure acceptable frequency performance under disturbances, especially in systems with high RES penetration. In addition, the spatial distribution of inertia is typically overlooked. In particular, the electrical distance between generation units and disturbance locations, which influences effective inertia contribution and frequency response, is seldom considered in the optimization process.

Motivated by these research gaps, this paper proposes a PINN framework to solve the OPF problem in RES–dominated power systems. The proposed model explicitly incorporates a location-aware inertia constraint that accounts for electrical distance by introducing the concept of system inertia strength. This constraint is integrated directly into the OPF formulation, while the governing physical equations of the power system are embedded within the neural network training process. By bridging steady-state optimal operation with inertia strength considerations within a unified learning-based framework, this work contributes toward the secure and efficient operation of future IBR-dominated power systems.

## II. Methodology

This section illustrates the mathematical modeling of the inertia-constrained OPF problem and the proposed PINN framework used to solve it. First, the OPF model and its operational constraints are described. Then, the architecture of the neural network and its physics-informed structure are introduced. Finally, the loss function formulation is presented.

### *A. RES-Integrated OPF*

The OPF problem aims to determine the most economical operating point of a power system while satisfying network constraints and operational limits. In this work, a DC Optimal Power Flow (DCOPF) framework is adopted and extended to explicitly incorporate RESs into the formulation. The extended DCOPF model is described by the following set of equations.

$$Min\left\{\sum_{i\in\Im} C_{con,i} + \sum_{i\in\Re} C_{res,i} + \sum_{i\in\Re} C_{res,i}^{HR}\right\} \tag{1}$$

$$C_{con,i} = \alpha_i P_{con,i}^2 + \beta_i P_{con,i} + \gamma_i ,\forall i \in \Im \tag{2}$$

$$C_{res,i} = \kappa_i P_{res,i}, \forall i \in \Re \tag{3}$$

$$C_{res,i}^{HR} = \kappa_i^{HR}\left(P_{res,i}^{\max} - P_{res,i}\right), \forall i \in \Re \tag{4}$$

$$\overline{B}\cdot\overline{\theta} = \overline{P_G} - \overline{P_D}, \theta_1 = 0 \tag{5}$$

$$P_{con,i}^{\min} \le P_{con,i} \le P_{con,i}^{\max}, \forall i \in \Im \tag{6}$$

$$P_{res,i}^{\min} \le P_{res,i} \le P_{res,i}^{\max}, \forall i \in \Re \tag{7}$$

$$\left|F_l\right| \le F_l^{\max}, l \in \Gamma \tag{8}$$

$$F_l = B_{l_1 l_2}\left(\theta_{l_1} - \theta_{l_2}\right), l \in \Gamma \tag{9}$$

The objective function of the OPF problem, which minimizes the total system operating cost, is shown in Equation (1). This total cost consists of three components: the generation cost of conventional units, the generation cost of RES units, and the headroom reservation cost of RESs. The headroom reservation cost of an RES represents the cost associated with operating below the available forecasted power in order to provide virtual inertia in the system. Equations (2)–(4) define these cost components in detail. Equation (2) represents the cost of conventional generation, (3) represents the cost of RES generation, and (4) represents the RES headroom reservation cost.

Equation (5) represents the DC load flow equation. Constraints (6) and (7) impose the minimum and maximum generation limits for conventional and RES units, respectively. Constraint (8) enforces transmission line capacity limits, ensuring that line flows do not exceed their thermal ratings. Finally, Equation (9) defines the calculation of line power flows.

### B. Inertia-Integrated OPF

System inertia plays a crucial role in maintaining frequency stability, particularly under high penetration of RESs. As conventional synchronous generation is progressively replaced by converter-interfaced resources, the overall system inertia decreases. In such low-inertia systems, frequency deviations following power imbalances become faster and more severe. Therefore, it is essential to explicitly consider inertia within the OPF framework to ensure that a minimum level of inertia is maintained in the system, thereby preventing excessive frequency excursions and potential system failure during disturbances.

Moreover, inertia is inherently location-dependent. Unlike many existing studies that treat inertia as a system-wide aggregated parameter, the distribution of inertia across the network affects frequency dynamics. The impact of inertia depends on its electrical proximity to disturbance locations and load centers. For instance, buses electrically closer to large synchronous generators typically experience stronger inertial support, while remote areas may be more vulnerable to frequency deviations. Therefore, neglecting the spatial distribution of inertia may lead to inaccurate assessments of system security [25].

For this reason, this paper incorporates a location-aware inertia constraint into the previously formulated RES-integrated OPF model. In this regard, a grid inertia strength metric is formulated and introduced in [25]. To calculate the inertia strength, a bus-weighting matrix is first defined as in (10) to capture the effect of electrical distance on inertia distribution across the network. This matrix enables a systematic representation of location-dependent inertia contributions, allowing the model to account for the spatial effectiveness of inertia support at different buses.

$$\overline{W} = -\left(\overline{B_{bb}}\right)^{-1} \cdot \overline{B_{bg}} \tag{10}$$

Where, the matrices $\overline{B_{bb}}$ and $\overline{B_{bg}}$ in (10) are obtained by partitioning the network susceptance matrix according to generator buses and non-generator buses, as in (11).

$$\overline{B} = \begin{bmatrix} \overline{B_{gg}} & \overline{B_{gb}} \\ \overline{B_{bg}} & \overline{B_{bb}} \end{bmatrix} \tag{11}$$

Given the bus-weighting matrix, the strength of the grid inertia could be calculated through (12) and (13).

$$\overline{H_B} = \overline{W} \cdot \overline{H_G} \tag{12}$$

$$H_{str} = \overline{I}_{H_B}^T \cdot \overline{H_B} \tag{13}$$

Therefore, the inertia strength could be achieved through the consideration of the power network and the electrical distance in between the buses.

For RES units to be able to contribute to system inertia, a certain amount of headroom is reserved for each unit. This reserved headroom enables the provision of virtual inertia through controlled power injection during frequency transients. The virtual inertia contribution of each RES unit is calculated using Equation (14), while the corresponding stored (or releasable) energy associated with the reserved headroom is computed in (15).

$$H_i^{HR} = \frac{E_i^{HR}}{\overline{I}_{P_D}^T \cdot \overline{P_D}}, \forall i \in \Re \tag{14}$$

$$E_i^{HR} = \left(P_{res,i}^{\max} - P_{res,i}\right)\tau_i^{HR}, \forall i \in \Re \tag{15}$$

Finally, to ensure the minimum strength of grid inertia (i.e., $H_{str}^{\min}$) in the system, the following constraint (16) could be considered, along with (12) to (15), in the OPF problem.

$$H_{str} \geq H_{str}^{\min} \tag{16}$$

### C. PINN Model

PINNs integrate physical equations into the process of training via the loss function, ensuring solutions satisfy physical constraints while reducing reliance on large datasets [26], [27]. While being applicable even to physical systems governed by differential equations, PINNs are also well suited to handling algebraic equations in power systems, such as power flow equations and operational constraints. Unlike conventional data-driven models, PINNs inherently enforce system physics and security constraints, making them particularly effective for OPF problems where feasibility is essential.

In the proposed framework, the PINN is used to approximate the solution mapping between system loading conditions and optimal generation dispatch while enforcing the load flow equations and the location-aware inertia constraint through the loss function.

The architecture of the proposed PINN follows a standard fully connected feedforward neural network structure [28], as illustrated in Fig. 1. The neural network is composed of an input layer, several hidden layers, and an output layer. Within each hidden layer, an affine mapping is applied, followed by a nonlinear activation function. For a neural network, the forward propagation can be expressed as:

$$Z_k = \sigma\left(w_k Z_{k-1} + b_k\right) \tag{17}$$

Where, $w_k$ and $b_k$ are the trainable weight matrices and bias vectors of layer $k$, $\sigma(\cdot)$ denotes the activation function, and $Z_k$ represents the input vector.

The input and output of the neural network are the demands at load buses (i.e., $\overline{P_D}$ ) and the predicted optimal dispatch at generator buses (i.e., $\overline{P_G}$ ), respectively. During training, the network parameters are optimized by minimizing a composite loss function that includes load flow equation, line flow limits, and the location-aware inertia requirement.

The choice of activation function plays an important role in training stability and convergence. Traditional activation functions such as rectified linear unit (ReLU) and tanh have been widely used in power system learning applications. However, ReLU may suffer from dead neuron problems due to zero gradients for negative inputs, and tanh may experience gradient saturation for large input magnitudes. Hence, in this work, the Sigmoid Linear Unit (SiLU), or the Swish activation function, is adopted. The SiLU function is defined as:

$$\sigma(z) = SiLU(z) = \frac{z}{1+e^{-z}} \tag{18}$$

By learning the mapping from load demands to economically optimal and inertia-secure dispatch, the proposed PINN offers a computationally efficient alternative to iterative solvers while maintaining physical feasibility and security constraints.

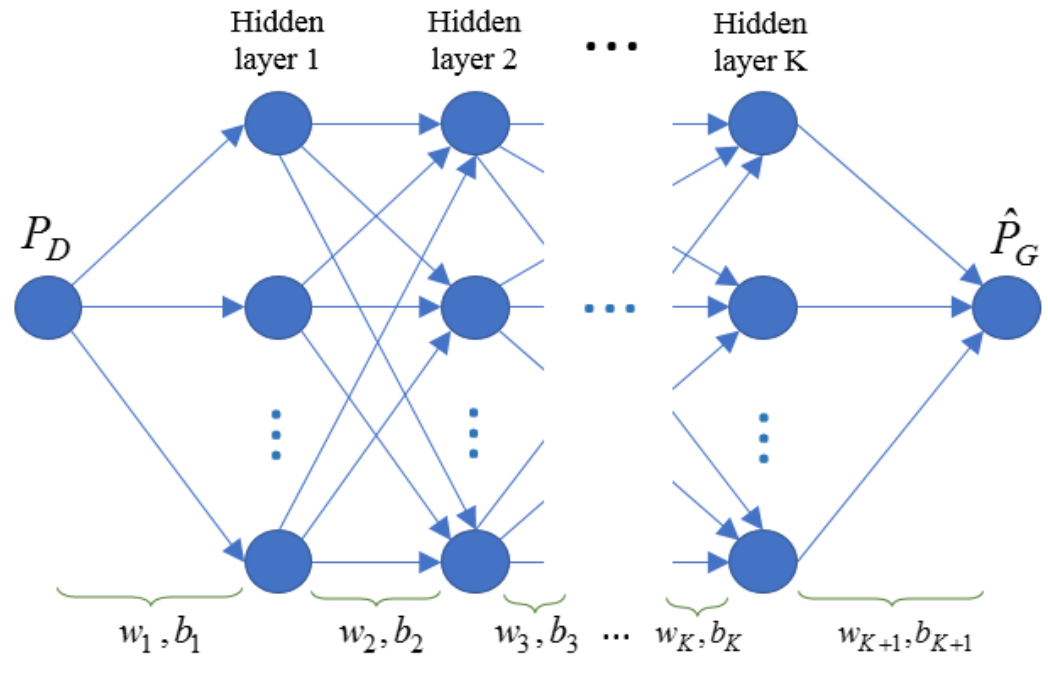

Fig. 1. The structure of the PINN

### D. Loss Function Formulation

The proposed PINN is trained by minimizing a composite loss function that integrates supervised learning with physical and operational constraints of the OPF problem. Instead of relying solely on data regression, the loss function embeds the power flow equations, transmission line limits, and system inertia requirements directly into the training objective. In this context, the total loss function is defined as:

$$L_{total} = \lambda_{data} L_{data} + \lambda_{DCLF} L_{DCLF} + \lambda_{flow} L_{flow} + \lambda_{iner} L_{iner} \quad (19)$$

Where, the coefficients $\lambda$ are positive weighting parameters that balance the relative importance of each component, and the loss terms $L$ are defined as follows.

$$L_{data} = \frac{1}{N}\sum_{j=1}^{N}\left\| \overline{\hat{P}_G}^{(j)} - \overline{P_G^*}^{(j)} \right\|_2^2 \quad (20)$$

$$L_{DCLF} = \frac{1}{N}\sum_{j=1}^{N}\left\| r^{(j)} \right\|_2^2 \quad (21)$$

$$L_{flow} = \frac{1}{N}\sum_{j=1}^{N}\sum_{l\in\Gamma}\left[ \mathrm{ReLU}\left( \left| F_l^{(j)} \right| - F_l^{\max} \right) \right]^2 \quad (22)$$

$$L_{iner} = \frac{1}{N}\sum_{j=1}^{N}\left[ \mathrm{ReLU}\left( H_{str}^{\min} - H_{str}^{(j)} \right) \right]^2 \quad (23)$$

In (20), the data loss enforces agreement between the neural network output and the reference OPF solutions, ensuring supervised learning accuracy. Equation (21) defines the physics-based power flow loss, where the residual of the DC power flow equations is penalized to guarantee network feasibility and power balance satisfaction. In this equation, the residual vector $r$ is calculated via (24) which is derived from the load flow equation (5). In this regard, the slack bus equation is excluded, and its corresponding residual component $r_0$ is not penalized.

$$\begin{bmatrix} r_0 \\ r \end{bmatrix} = \overline{B} \cdot \overline{\theta} - \left( \overline{P_G} - \overline{P_D} \right) \quad (24)$$

In (22), the transmission congestion loss penalizes branch flow violations, enforcing thermal limits without affecting feasible operating points. Finally, Equation (23) introduces the inertia constraint loss, which imposes a minimum inertia strength requirement. The strength of inertia is computed in the system for each data, and the inadequate values are penalized. It is noteworthy that in (22) and (23), ReLU function is the rectified linear unit, mathematically represented as follows:

$$\mathrm{ReLU}(z) = \max\{0, z\} \quad (25)$$

Furthermore, the generation limit constraints—constraints (6) and (7)—are enforced directly at the neural network output layer using a sigmoid-based scaling transformation. To do so, the unconstrained network output is mapped into the feasible interval $\left[ P_{G,i}^{\min}, P_{G,i}^{\max} \right]$ via:

$$P_{G,i} = P_{G,i}^{\min} + \left( P_{G,i}^{\max} - P_{G,i}^{\min} \right) \sigma'\left( z_i \right), \forall i \in \left( \Im \cup \Re \right) \quad (26)$$

Where,

$$\sigma'(z) = \left( 1 + e^{-z} \right)^{-1} \quad (27)$$

This approach embeds the inequality constraints into the network architecture, eliminating the need for additional penalty terms and ensuring feasibility by construction.

## III. Case Study

To evaluate the effectiveness of the proposed method, the well-known IEEE 39-bus system is adopted as the benchmark network, as in [29]. The total installed generation capacity of the modified system is assumed to be 6 GW. In this configuration, five conventional generation units are located at buses 1, 5, 10, 15, and 20, while five RES units are placed at buses 31, 33, 35, 37, and 39. These units are labeled sequentially as Units 1 to 10, respectively. The system includes 19 load buses, namely buses 3, 4, 7, 8, 12, 15, 16, 18, 20, 21, 23, 24, 25, 26, 27, 28, 29, 31, and 39.

Each conventional generation unit is assumed to have an inertia constant of 0.2 seconds. In the event of a power imbalance, each RES unit is assumed to provide frequency support for up to 10 seconds. The minimum required inertia strength of the system is set to 5 seconds to ensure acceptable frequency stability.

A PINN is employed to learn the optimal generation dispatch corresponding to different loading conditions. The network structure includes three hidden layers with 128 neurons in each layer. The loss function incorporates weighted components with $\lambda_{data} = 1$ and $\lambda_{DCLF} = \lambda_{flow} = \lambda_{iner} = 0.1$ to properly balance the physical constraints and data-driven terms during training.

The dataset used for training and testing is generated in MATLAB. For input generation, each load bus is initially assigned a random demand between 0 and 0.1 p.u. The loads are then proportionally scaled so that the total system demand lies in range $[0.95, 1.05]$ p.u., thereby simulating realistic variations around nominal loading conditions. For every generated load profile, the corresponding optimal generation dispatch is computed using MATLAB's optimization toolbox. These optimal solutions serve as the ground truth outputs for training the PINN. In total, 1000 samples are generated for training and 10,000 samples are generated for testing.

To assess the predictive accuracy of the trained network, four performance metrics are considered for each individual generation unit as well as for the total system generation. The first is the Mean Absolute Error (MAE), which quantifies the average absolute discrepancy between the predicted and optimal generation outputs, thereby indicating the overall magnitude of the errors. The second is the Root Mean Square Error (RMSE), which places greater emphasis on larger deviations and is therefore more sensitive to significant prediction errors. The third metric is the Mean Absolute Percentage Error (MAPE), which measures the average relative error in percentage terms. The fourth metric is the Maximum Absolute Error (MaxAE), which represents the largest absolute deviation observed among all samples and therefore indicates the worst-case prediction error.

These metrics are computed separately for the training and testing datasets and are illustrated in Fig. 2 and Fig. 3, respectively.

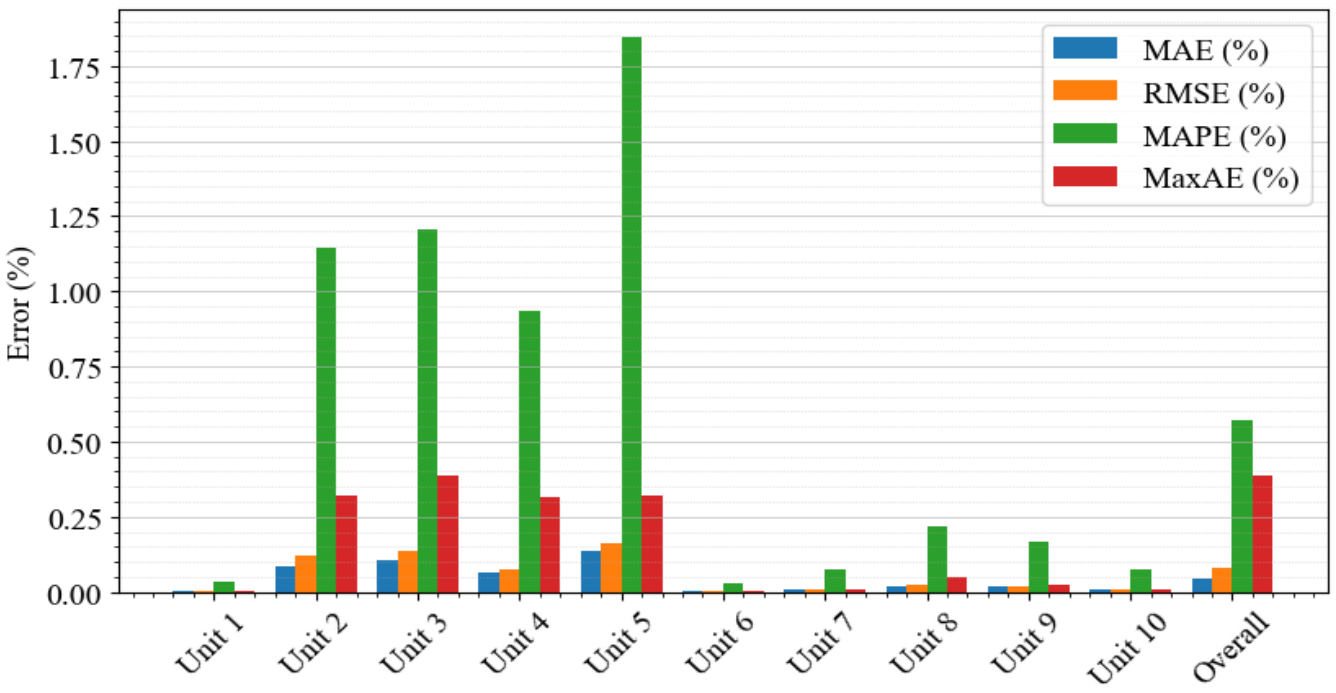

Fig 2. performance metrics for the training dataset.

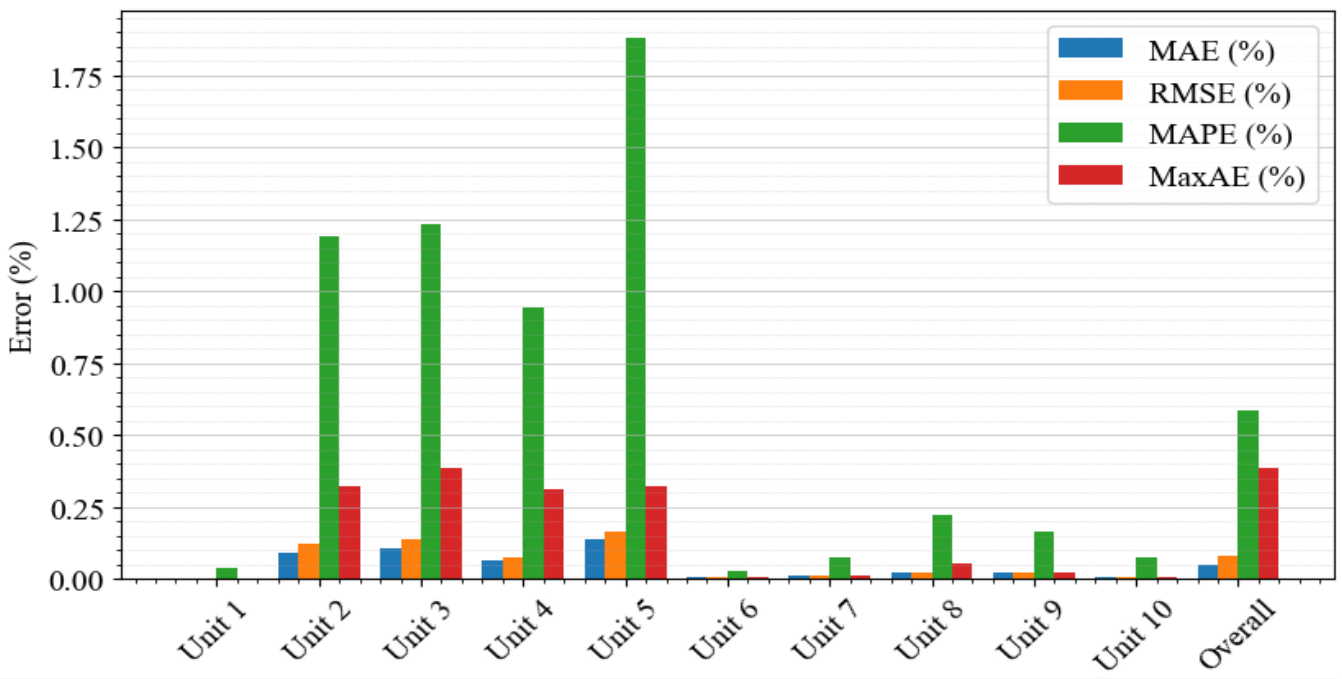

Fig 3. performance metrics for the testing dataset.

The overall calculated metrics are represented in Table 1 in exact values for both the training and testing datasets. As shown in the table, the error values for the training and testing datasets are very close to each other, indicating good generalization capability and the absence of overfitting. In particular, the MAE for both datasets is approximately 0.045% of the total system capacity.

To further validate the model performance, several representative generation units are selected for detailed comparison between the optimal and predicted outputs. Specifically, Units 3, 4, 5, and 8 are examined. For this purpose, 50 randomly selected load profiles from both the training and testing datasets are considered. The corresponding optimal generation values and the PINN predictions are depicted in Fig. 4 for the training dataset and Fig. 5 for the testing dataset. The results show a strong agreement between the predicted and optimal values across different operating conditions. The close alignment of the curves confirms the ability of the proposed PINN framework to accurately approximate the optimal dispatch mapping under varying load scenarios.

In addition, the learning curve of the PINN is presented in Fig. 6 Note that the curve is plotted using a moving average to smooth out sharp fluctuations, thereby providing a clearer visualization of the training trend and convergence behavior.

Table 1. Error metrics for training and testing datasets.

| Error metrics | MAE (%) | RMSE (%) | MAPE (%) | MaxAE (%) |
|---|---|---|---|---|
| Training data set | 0.045535 | 0.080273 | 0.573 | 0.386 |
| Testing data set | 0.046426 | 0.081919 | 0.585 | 0.386 |

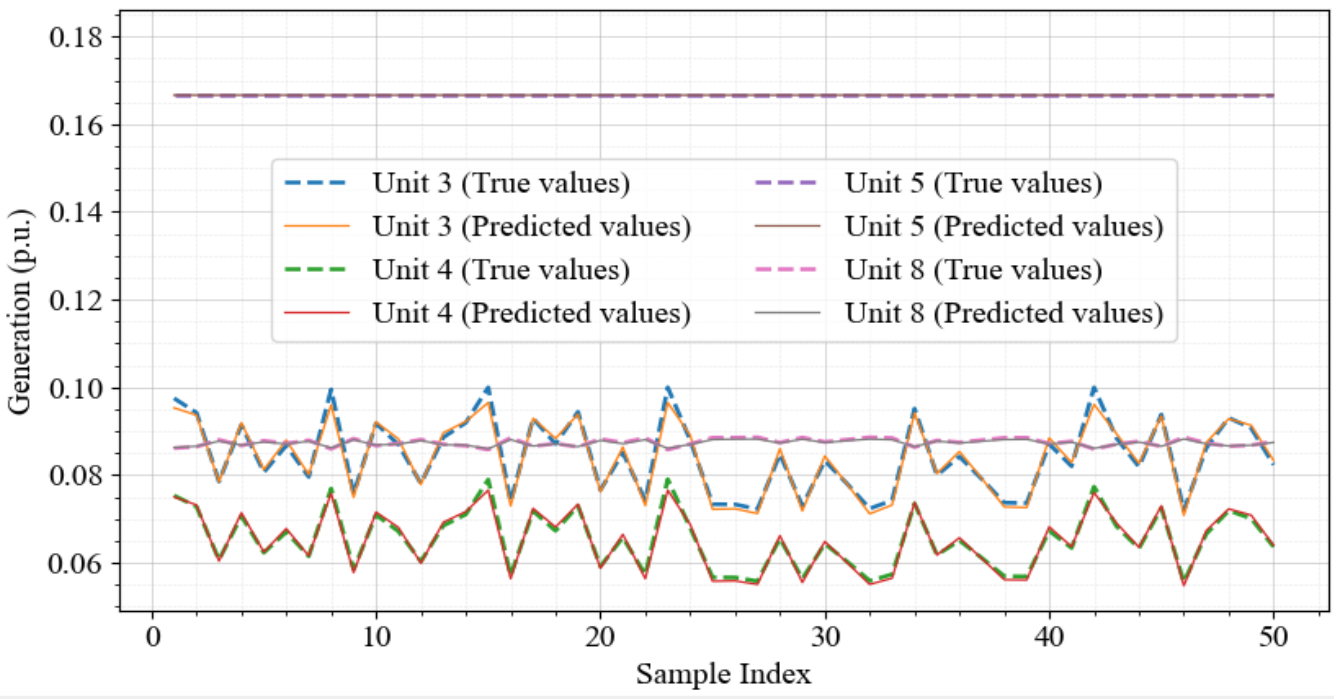

Fig 4. True and predicted values of selected units for training dataset.

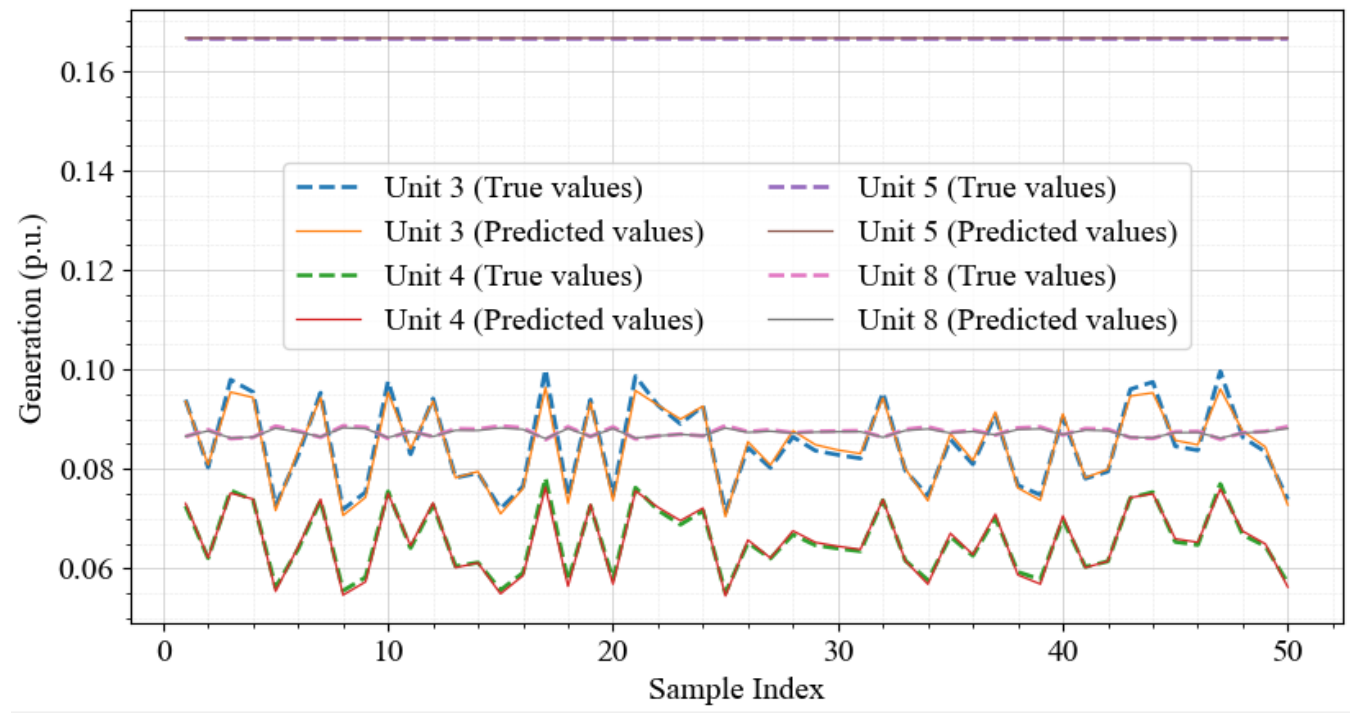

Fig 5. True and predicted values of selected units for testing dataset.

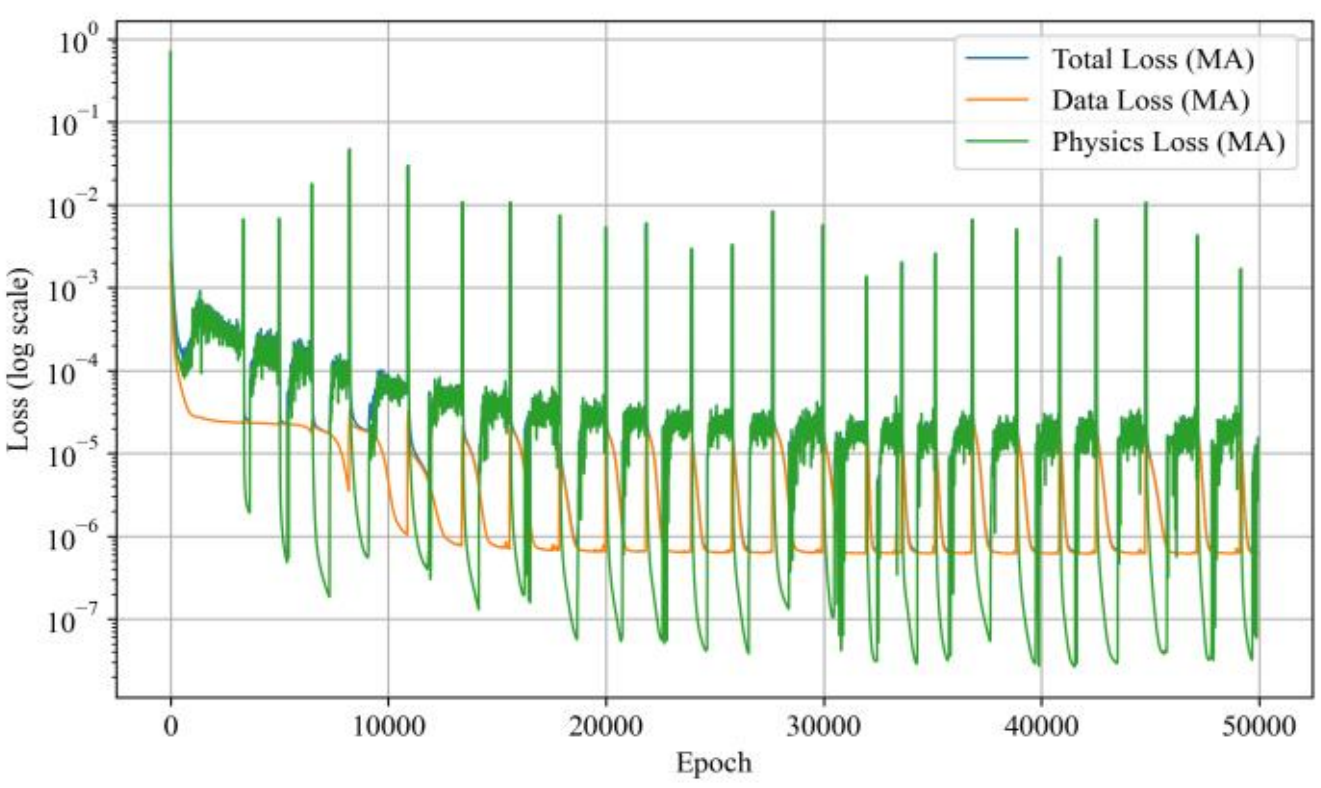

Fig 6. The PINN learning curve.

Fig. 7 illustrates the variation of MAE with respect to different training dataset sizes, serving as a data-size sensitivity analysis of the proposed PINN framework. As the amount of training data increases, the MAE on the training set shows a slight increase, while the MAE on the testing set decreases. This behavior indicates that, although the model fits the training data less tightly as the dataset becomes larger and more diverse, its generalization capability improves significantly on unseen data.

Furthermore, Fig. 8 presents the MAE values obtained under different physics-factor settings. Here, the term physics factor refers collectively to the weighting coefficients $\lambda_{DCLF}$ , $\lambda_{flow}$ , and $\lambda_{iner}$ used in the physics-informed loss function. As shown in the figure, the factor value of 0.1 yields the lowest MAE among all tested cases, indicating the best balance between data-driven learning and physics-based constraint enforcement, thereby being adopted in this paper.

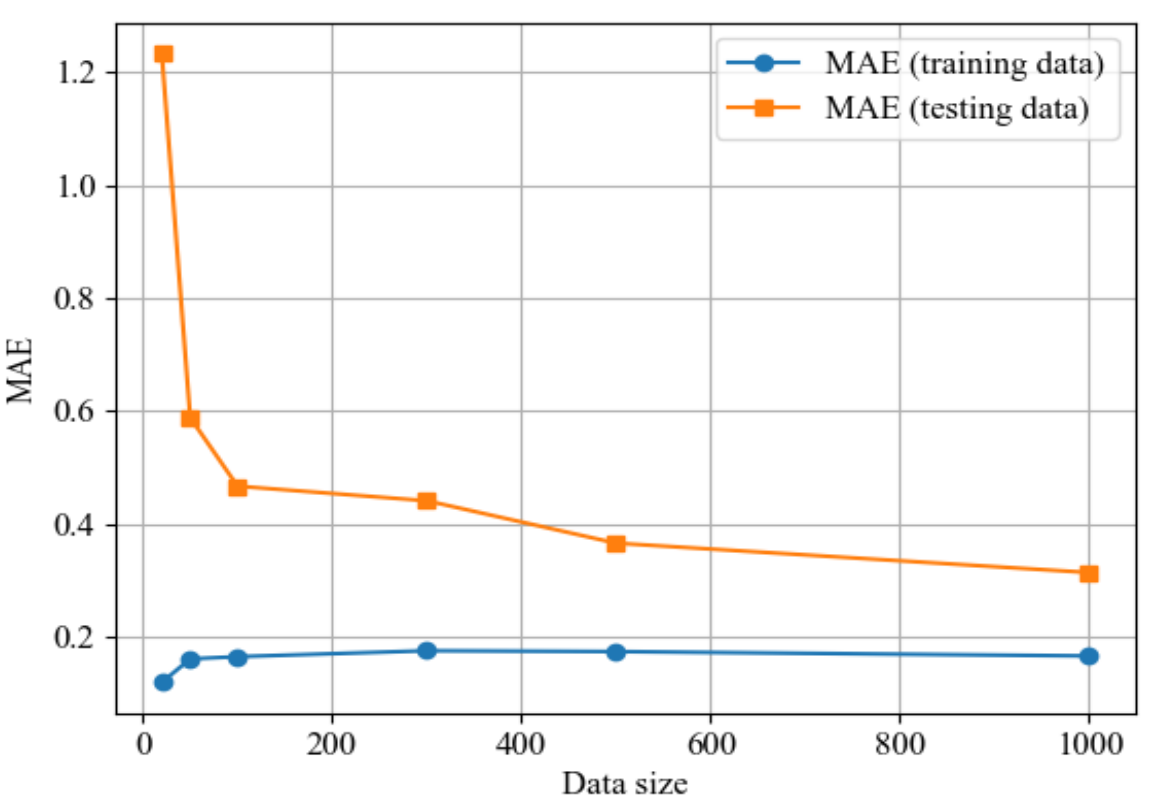

Fig 7. Comparison of MAE values for different data sizes.

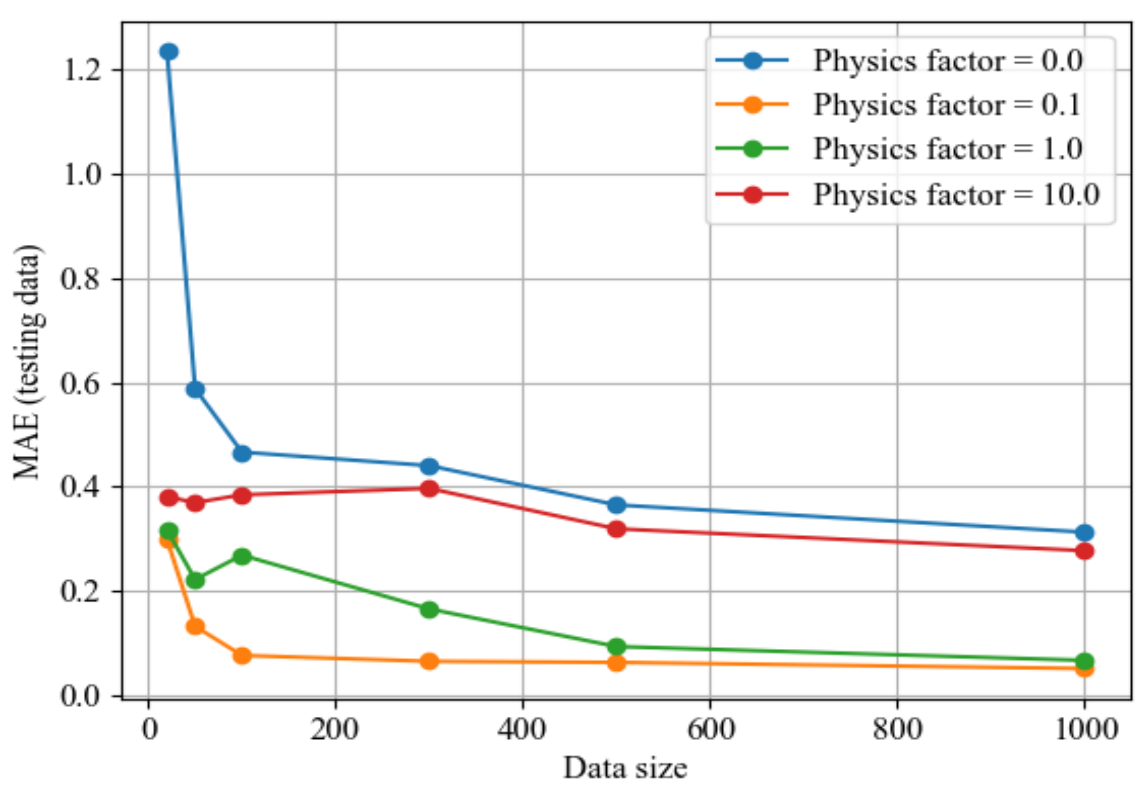

Fig 8. Comparison of MAE values for different physic factors.

## IV. Conclusion

This paper introduced a PINN-based framework for solving the OPF problem in renewable-dominated, low-inertia systems with an explicit location-aware inertia constraint. By embedding power flow physics into the learning process and incorporating system inertia strength into the optimization, the proposed method unifies economic operation and frequency security. Numerical results on a 6 GW test system confirm the high accuracy and strong generalization capability of the proposed approach. The mean absolute error for both the training and testing datasets is approximately 0.045% of the total system capacity, indicating only negligible deviation from the reference OPF solutions. These findings demonstrate that the proposed framework provides a computationally efficient, scalable, and physically consistent solution for inertia-aware OPF in future IBR-dominated power systems.

## V. Acknowledgement

This work was supported by the European Union's Horizon Research and Innovation Programme under the agreement No. 101120657 through the ENFIELD project (European Lighthouse to Manifest Trustworthy and Green AI) in collaboration with Know Center. Know Center is part of the COMET programme, which is coordinated by the Austrian Research Promotion Agency (FFG) and supported by Austrian federal and regional funding bodies.